\documentclass[journal]{IEEEtai}

\usepackage[colorlinks,urlcolor=blue,linkcolor=blue,citecolor=blue]{hyperref}

\usepackage{color,array}

\usepackage{graphicx}
\usepackage{amsmath}
\usepackage{algorithm}
\usepackage{algorithmicx}
\usepackage{algpseudocode}
\usepackage{cite}

\begin{document}

\title{Large Language Model Guided Self-Debugging Code Generation
\thanks{This work is funded under the agreement with the ACT Government, Future Jobs Fund - Open Source Institute(OpenSI) - R01553; and NetApp Technology Alliance Agreement with OpenSI - R01657.
Additionally, this research was supported by the Australian Government through the Department of Education’s National Industry PhD Program (project 36337). 
The views expressed herein are those of the authors and are not necessarily those of the Australian Government or the Department of Education.}
}

\author{Muntasir Adnan\IEEEauthorrefmark{1},
Zhiwei Xu\IEEEauthorrefmark{2},
and Carlos C. N. Kuhn\IEEEauthorrefmark{1}
\thanks{\IEEEauthorrefmark{1}Open Source Institute, Faculty of Science and Technology, University of Canberra, Canberra, Australia.}
\thanks{\IEEEauthorrefmark{2}School of Computing and Information Systems, The University of Melbourne, Australia.}
\thanks{Corresponding author: Muntasir Adnan (Adnan.adnan@canberra.edu.au).}
}

\maketitle

\begin{abstract}
Automated code generation is gaining significant importance in intelligent computer programming and system deployment. 
However, current approaches often face challenges in computational efficiency and lack robust mechanisms for code parsing and error correction. 
In this work, we propose a novel framework, PyCapsule, with a simple yet effective two-agent pipeline and efficient self-debugging modules for Python code generation. 
PyCapsule features sophisticated prompt inference, iterative error handling, and case testing, ensuring high generation stability, safety, and correctness. 
Empirically, PyCapsule achieves up to 5.7\% improvement of success rate on HumanEval, 14.6\% on HumanEval-ET, and up to 24.5\% on BigCodeBench compared to state-of-the-art methods. 
Notably, PyCapsule enables the open-source Qwen2.5-Coder-Instruct 7B model to surpass the standalone performance of GPT-4 and GPT-3.5 on HumanEval and MBPP while requiring an average of fewer than 2 API calls per problem compared to up to 17 in competing frameworks. 
PyCapsule demonstrates broader impacts on advancing lightweight and efficient code generation, making competitive automated programming attainable with cost-effective, locally deployable open-source models.
\end{abstract}

\begin{IEEEImpStatement}
Automated code generation is an increasingly vital capability in modern software development, reducing the burden on human developers and accelerating deployment across industry and research. 
However, existing frameworks often demand substantial computational resources, limiting accessibility for organisations without large-scale infrastructure. 
PyCapsule addresses this directly through a lightweight two-agent architecture that delivers state-of-the-art performance while requiring an average of fewer than two API calls per problem, compared to up to seventeen in competing frameworks. 
By enabling an open-source 7-billion-parameter model to surpass the standalone performance of much larger proprietary systems such as GPT-4 on standard benchmarks, PyCapsule demonstrates that high-quality automated programming is achievable without resource-intensive infrastructure.
As organisations increasingly turn to open-source models to control costs and retain governance over proprietary code, PyCapsule offers a practical route to competitive code generation that runs on accessible, locally deployable infrastructure.
\end{IEEEImpStatement}


\section{Introduction}
\label{intro}
The evolution of generative artificial intelligence has profoundly influenced multiple domains, notably in automated code generation.
In particular, Large Language Models (LLMs) have demonstrated remarkable proficiency in code generation from natural language prompts, streamlining software development processes and enhancing productivity~\cite{emperical_study_soft_eg, emp_study_critical, palm10.5555/3648699.3648939}.
This transformative potential has sparked significant research in this field~\cite{islam2024mapcodermultiagentcodegeneration, ddi, ALPHA}.
However, several critical challenges emerge as these systems scale to handle increasingly complex programming tasks.
Ensuring reliability, accuracy, and autonomy in code generation remains critical but challenging, particularly in error detection and debugging~\cite{llm_se_problems, dou2024whatswrongcodegenerated, imperfect_code_gen}. 
Recent studies have introduced frameworks consisting of multiple LLM-based agents to emulate human problem-solving approaches in code generation. 
For instance, MapCoder ~\cite{islam2024mapcodermultiagentcodegeneration}, Debugging Decay Index~\cite{ddi}, and LDB ~\cite{zhong2024debuglikehumanlarge} employ agents dedicated to retrieval, planning, coding, and debugging to enhance code quality. 
While these approaches have shown remarkable improvements, they often require substantial computational resources and complex agent coordination.
This poses challenges for real-world deployment and scalability.
Moreover, many high-performing frameworks~\cite{islam2024mapcodermultiagentcodegeneration, hong2024metagpt} rely on large proprietary models, raising concerns around cost, control, and data privacy, and contributing to the growing interest in locally deployable open-source solutions for code generation.

To address these concerns, we propose a novel framework, PyCapsule, that enhances the efficiency and accuracy of LLM-based Python code generation through a two-agent architecture, demonstrating that locally deployed open-source models can approach and often exceed the performance of frameworks relying on large proprietary systems.
This simple yet effective two-agent architecture maintains a clear separation of responsibilities without the overhead of additional agents, thereby minimising computational consumption.
PyCapsule employs a \textit{programmer agent} responsible for code generation and debugging, interacting with an \textit{executor agent} that handles code validation, case testing, and error analysis.
The framework's effectiveness is enhanced by three specialised modules: an \textit{error handler} that refines the debugging feedback, an \textit{example call detector} that prevents runtime issues, and a \textit{function signature converter} that provides a uniform template for the function description. 
This structure enables PyCapsule to achieve high performance on widely used datasets, including HumanEval~\cite{codex}, HumanEval-ET~\cite{dong2023codescore}, Mostly Basic Python Programming (MBPP)~\cite{austin2021program}, MBPP-ET~\cite{dong2023codescore}, and BigCodeBench~\cite{zhuo2024bigcodebench} with a high computational efficiency.
The main contributions of our work are summarised as follows --- 
\begin{itemize}
    \item Our two-agent architecture implemented in PyCapsule demonstrates that complex coordination between multiple agents can be replaced by well-structured language prompts and specialised modules without sacrificing accuracy, requiring fewer computational resources compared to other self-debugging frameworks.

    \item We introduce an efficient self-debugging mechanism by combining the programmer agent and executor agent with three specialised modules: an \textit{error handler}, an \textit{example call detector}, and a \textit{function signature converter}. 
    These achieve up to 5.7\% accuracy improvement on HumanEval, 
    14.6\% on HumanEval-ET, and up to 24.5\% on BigCodeBench, while enabling the open-source Qwen2.5-Coder-Instruct 7B model to surpass the standalone performance of GPT-3.5 and GPT-4 on HumanEval.

    \item Through an ablation study and comprehensive evaluation across multiple datasets and models, we validate the independent contribution of each module in PyCapsule's architecture, and corroborate the exponential decay in iterative debugging effectiveness~\cite{ddi} as a consistent phenomenon across model families and scales.
\end{itemize}

\section{Background}
\label{lit}
Automated high-quality code generation has evolved in several key directions related to prompt engineering~\cite{ai_literacy_p_e, Wei2022ChainOT}, multi-agent systems~\cite{hong2024metagpt, wooldridge1995intelligent, approach_to_ai_agents}, and iterative debugging with runtime feedback mechanisms~\cite{generative_agents, zhong2024debuglikehumanlarge, chen2023teachinglargelanguagemodels, islam2024mapcodermultiagentcodegeneration, software_testing}.
While LLM has demonstrated increasing capability in code generation, the code quality and correctness can be further enhanced by using intelligent debugging strategies.
Previous works in self-debugging~\cite{Huang2023AnES, chen2023teachinglargelanguagemodels} show their capability in Automated Program Repair (APR) by identifying and fixing significant code flaws and bugs.
Researchers have shown that ``large language models of code"~\cite{Huang2023AnES} often struggle to directly fix bugs due to the lack of task-specific fine-tuning. 
While fine-tuned models can often outperform state-of-the-art (SOTA) APR tools~\cite{lesstraining, vulrepair, repairisnearly}, they still suffer from loss of pre-trained knowledge~\cite{Xia2022LessTM}, inferior long-sequence handling ability, high resource constraints, long-tail problems, and multi-hunk error fixing~\cite{Huang2023AnES}.

MapCoder~\cite{islam2024mapcodermultiagentcodegeneration} presents a multi-agent framework designed to emulate human-like programming practices. 
It employs four agents for retrieval, planning, coding, and debugging to perform iterative code generation and debugging.
MetaGPT~\cite{hong2024metagpt} introduces a meta-programming framework that encodes Standardised Operating Procedures (SOPs) into prompt sequences, assigning distinct roles such as product manager, architect, and engineer to separate agents in an assembly-line paradigm. 
The ``Debug Like a Human" framework~\cite{zhong2024debuglikehumanlarge} introduces Large Language Model Debugger (LDB), which enhances the code quality by leveraging runtime execution information. 
LDB iteratively segments code, tracks intermediate variable states, and performs step-by-step verification like breakpoints.

Some approaches have employed prompt engineering to improve the quality of code generation and debugging.
For instance,  CodeCoT~\cite{huang2024codecottacklingcodesyntax} leverages iterative refinement through Chain of Thought (CoT)~\cite{Wei2022ChainOT} and logical reasoning. 
However, its capability is limited to refining syntax errors.

While these approaches have advanced automated code generation, they share several notable limitations.
Frameworks such as MapCoder, MetaGPT, and LDB incur substantial computational overhead, requiring complex agent coordination and high token consumption, which scales poorly with problem complexity.
Additionally, they often struggle to fully leverage error messages, which impairs their adaptability in nuanced debugging scenarios and can result in suboptimal code-generation performance. 
Beyond this, relying on complementary test case generation to compensate for evaluation gaps further increases computational demands and may reduce reliability~\cite{software_testing, li2024largelanguagemodelstest}, potentially necessitating more sophisticated agent architectures.

To address these challenges, we introduce PyCapsule as a streamlined, modular architecture with robust and optimised supporting modules. 
This substantially reduces framework complexity while facilitating more efficient, reliable and accurate code generation.

\section{Methodology}
\label{method}
PyCapsule employs two distinct types of agents, as defined in the agent-based computing literature~\cite{Nwana_1996, roadmap} --- smart agents and collaborative agents. 
The system includes two such agents --- a \textit{programmer agent} and an \textit{executor agent}, that embody these types respectively.
Following the agent typology in~\cite{Nwana_1996}, these agents are classified based on their emphasis on three key attributes: autonomy, cooperation, and learning.

\begin{figure}[!h]
    \centering
    \includegraphics[width=\linewidth, height = 5cm]{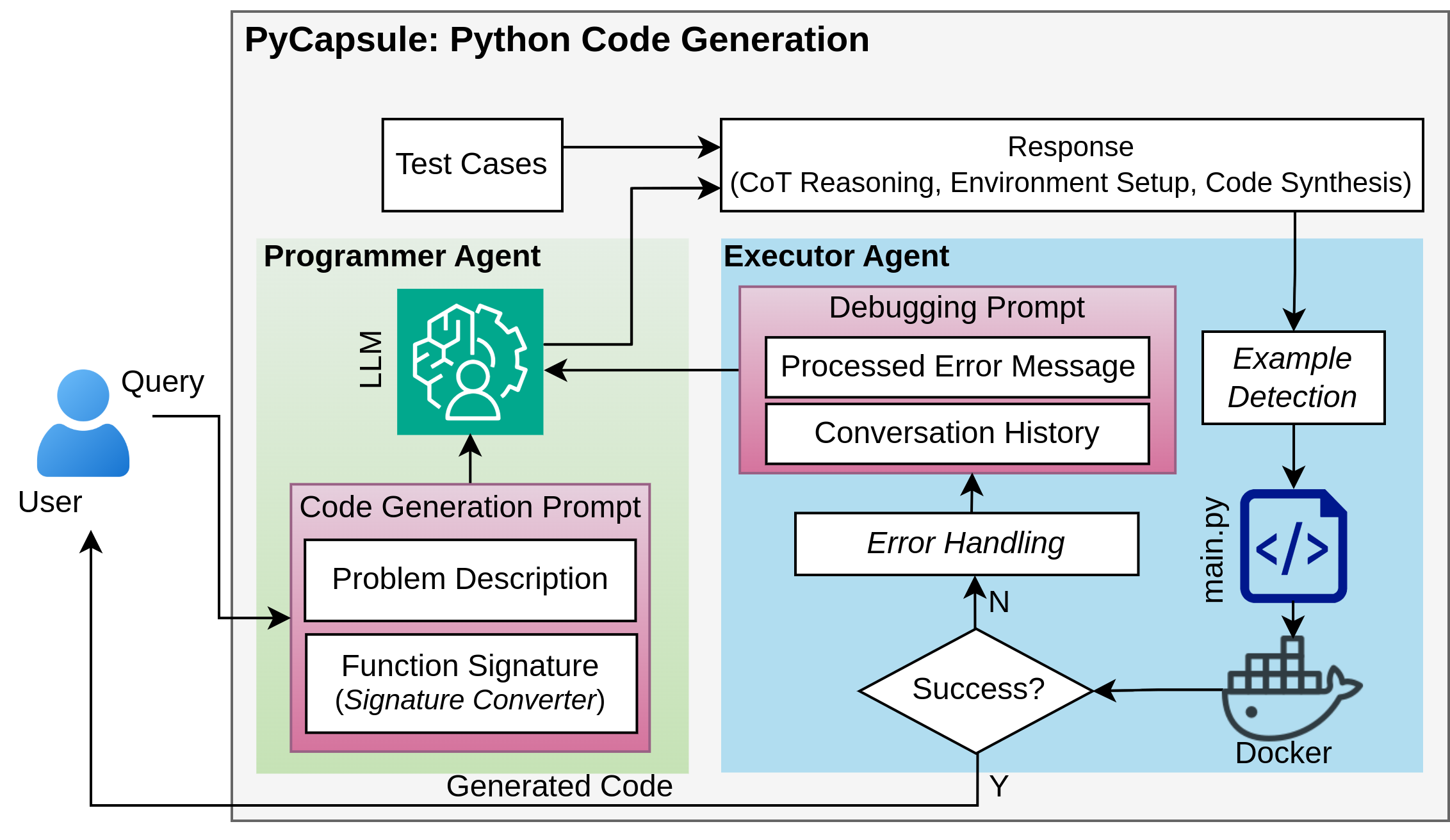}
    \caption{PyCapsule framework for Python code generation. The framework comprises iterative code generation, debugging, error handling, and code execution within a Docker container.}
    \label{fig: flowchart}
\end{figure}

\subsection{Programmer and Executor Agents}
\label{pycapsule}
The programmer agent functions as a smart agent, emphasising all three attributes through its LLM-based reasoning --- autonomous code-generation decisions, cooperative debugging interactions, and performance improvements across debugging iterations. 

The executor agent operates as a collaborative agent.
While implemented with a Docker container and supporting modules rather than an LLM, it meets the essential criteria of autonomy (operating without direct human intervention) and cooperation (communicating execution results and coordinating with the programmer agent). 
The executor agent acts as an autonomous validator that continuously monitors and responds to code execution outcomes, makes independent decisions about success or failure, and cooperates with the programmer agent by providing structured feedback to facilitate debugging.
When execution succeeds, it forwards the validated code to the user; otherwise, it generates detailed diagnostic reports and initiates a real-time feedback loop with the programmer agent for self-debugging. 
This interaction exemplifies the social ability identified in~\cite{roadmap, hong2024metagpt} as necessary for agent cooperation.

The programmer agent employs tailored system prompts with two programming modes --- \textit{generation mode} and \textit{fix mode}. 
These prompts, detailed in Appendix~\ref{appendix: prompt}, define the programmer agent's persona~\cite{role-play}, provide task-specific background, and simplify code extraction~\cite{li2025longcontext}.
In the \textit{generation mode}, the programmer agent uses CoT~\cite{Wei2022ChainOT} to analyse the problem description and devise a structured plan to generate executable code.
In the fix mode, the programmer agent attempts to debug code using refined error information from the previous execution attempt.
This fix-mode prompt is supported by two key information elements: conversation history and error messages processed by the \textit{error-handling module}. 
For conversation history management, we retain only the most recent problem-solution pair, consisting of the original problem description and the prior solution attempt.
This single pair defines the context for the current debugging attempt.
The choice to retain only the most recent aligns with Markov Decision Process (MDP), where the agent's decision depends only on the current state~\cite{Sutton1998}.
Furthermore, this design addresses a known challenge in LLM performance degradation as context length increases, attributed to attention dilution and information retrieval difficulties in longer contexts~\cite{li2025longcontext}. 
Empirical validation of this design choice is provided in Section~\ref{ablation}.

The Docker container of the executor agent uses a predefined shell script as its entry point, which facilitates environment setup and code execution. 
While this static script is fixed as a part of the executor agent, the programmer agent generates the solution code and any required dependency specifications for each problem. 
This separation allows for a consistent execution environment while enabling problem-specific implementations.
Upon executing the generated solution, the container returns either a success or a failure status. 
The success status indicates that the generated code has passed all test cases.
Alternatively, the failure status triggers the \textit{error-handling module} to relay diagnostics to the programmer agent, which then enters into fix mode for iterative self-debugging.

\subsection{Supportive Modules}
\label{aux}
Existing approaches like MapCoder~\cite{islam2024mapcodermultiagentcodegeneration} and LDB~\cite{zhong2024debuglikehumanlarge} rely heavily on LLM agents for code generation and validation.
However, this requires a high resource consumption.
In contrast, PyCapsule employs three specialised modules to handle each task deterministically --- a signature converter, an example call detector, and an error-handling module.

The \textit{signature converter} enhances code generation stability by extracting structured function signatures from problem descriptions.
This module generates a function name, a robust signature, and an example call using the first available test case for a given problem without revealing the test results, thereby avoiding extra LLM calls for function signature generation.

The \textit{example call detector} ensures the safety of code execution by removing any internal example function calls embedded in the generated code, which would otherwise execute unintentionally during test case evaluation.
This can result in unintended side effects, including infinite loops that exceed the maximum runtime limit and bypass runtime safety mechanisms. 
This deterministic module eliminates the need for a dedicated LLM agent to perform code-execution safety checks, reducing computational overhead.

The \textit{error-handling} module refines error messages returned by the executor agent, producing concise and contextually relevant natural language feedback. 
It identifies the error type, filters out irrelevant traceback information, and resolves critical issues, such as verbose recursion errors that can exceed the programmer agent's context length.
This systematic approach reduces token consumption during self-debugging iterations compared to passing raw, unprocessed error messages directly to the programmer agent.

These modules collectively enhance the code generation stability, safety, and self-debugging efficiency while reducing the computational overhead.

\subsection{Code Generation Workflow}
\label{workflow}
PyCapsule consists of three phases: code generation, execution, and self-debugging, shown in Figure~\ref{fig: flowchart}. 
This pipeline integrates the programmer and executor agents with the self-debugging mechanism to ensure reliable, efficient, and accurate code generation.

The workflow starts with the programmer agent receiving a problem description, typically supplemented by a function signature from the dataset or the signature converter module.
Subsequently, the agent generates a response comprising CoT reasoning, synthesised code and environment setup instructions.
This response is parsed to extract the synthesised code and environment setup instructions, the latter of which are used to automatically produce a \texttt{requirements.txt} file containing all required Python libraries.
The extracted code is further processed through the example call detector module to remove unintended example calls.
The final processed code is saved in \texttt{main.py}, alongside the dataset test cases, which are appended solely for evaluation purposes and are never exposed to the programmer agent.
To ensure execution safety, a maximum runtime timer is enforced to terminate any non-terminating execution, including infinite loops and blocking operations such as server initialisation within test functions, and to raise a timeout error if the execution limit is exceeded.

At the execution phase, the executor agent sets the Python environment using the requirements from \texttt{requirements.txt} and runs \texttt{main.py} in the Docker container.
The resulting execution status, either success or failure, determines whether the programmer agent's fix mode is invoked. 
If execution fails, error messages returned by the executor agent are refined by the error-handling module and passed to the programmer agent, which enters fix-mode for iterative code correction.
Building on the exponential decay in debugging effectiveness established in prior work~\cite{ddi} and further corroborated by our own analysis (see Section~\ref{result}), the programmer agent is limited to a maximum of five self-debugging attempts, beyond which returns are negligible.
Each self-debugging attempt is informed by the execution feedback from the previous iteration. 
The pseudo-code for the pipeline is provided in Appendix~\ref{appendix: pseudo}.

\section{Experiments}
\label{experiments}
All experiments involving locally deployed models were conducted on a single NVIDIA GeForce RTX 3090 (24 GB), while API-based models were accessed via their respective inference endpoints. 
We evaluate performance using success rate, defined as the ratio of correctly resolved problems to the total number of problems. 
Our code will be released upon publication.

\subsection{Dataset Overview}\label{dataset}
We used five benchmark datasets to evaluate PyCapsule --- HumanEval~\cite{codex}, HumanEval-ET~\cite{dong2023codescore}, MBPP~\cite{austin2021program}, MBPP-ET~\cite{dong2023codescore}, and BigCodeBench~\cite{zhuo2024bigcodebench},
where the ``ET" variants provide more test cases and enriched problem descriptions.
These datasets encompass diverse programming challenges.

HumanEval~\cite{codex}  consists of 164 human-curated programming problems covering string manipulation, arithmetic, and basic data structures, each comprising a natural language description, a function signature, and test cases.
MBPP~\cite{austin2021program} includes 974 programming problems spanning algorithms, data structures, and general-purpose tasks, providing natural language descriptions and test cases, which poses an additional challenge for code generation stability. 
BigCodeBench~\cite{zhuo2024bigcodebench} includes two difficulty splits --- full (1,140 problems) and hard, covering diverse programming tasks involving complex function calls, library usage, and real-world programming scenarios. 
The benchmark provides two prompt styles: an instruct split presenting each problem as a concise natural language instruction, and a complete split providing a detailed function signature and docstring. 
We evaluate PyCapsule on the full set, constructing a unified prompt by combining the natural language instruction from the instruct split with the function signature from the complete split, ensuring consistent input formatting across all problems.

\subsection{Results and Analysis}
\label{result}
\begin{table}[t]
\caption{Success Rate Comparison on Three LLMs. 
Mean and standard deviation for PyCapsule are reported over three independent runs. 
Results for all other methods are taken from their official reports.
``Direct" refers to the zero-shot setting.}
\label{table_acc}
\tablefont%
\setlength{\tabcolsep}{3pt}
\begin{tabular*}{21pc}{@{}|p{62pt}<{\centering}
                          |p{42pt}<{\centering}
                          |p{46pt}<{\centering}
                          |p{34pt}<{\centering}
                          |p{34pt}<{\centering}|@{}}
\hline
\textbf{Method} & 
\textbf{HumanEval} & 
\textbf{HumanEval-ET} & 
\textbf{MBPP} & 
\textbf{MBPP-ET} \\
\hline
\multicolumn{5}{|c|}{\textit{GPT-4-Preview-1106}} \\
\hline
Direct~\cite{clarify}
    & 78.86 & 70.73 
    & 81.1~\cite{islam2024mapcodermultiagentcodegeneration} & 51.52 \\
CoT~\cite{islam2024mapcodermultiagentcodegeneration}
    & 89.0 & -- & 82.4 & -- \\
Self-Planning~\cite{islam2024mapcodermultiagentcodegeneration}
    & 85.4 & -- & 75.8 & -- \\
MetaGPT~\cite{hong2024metagpt}
    & 85.9 & -- & 87.7 & -- \\
MapCoder~\cite{islam2024mapcodermultiagentcodegeneration}
    & 93.9 & 82.9 & 82.1 & 57.7 \\
\textbf{PyCapsule (ours)}
    & \textbf{96.5$\pm$0.7} 
    & \textbf{96.3$\pm$1.2} 
    & \textbf{88.2$\pm$0.3} 
    & \textbf{73.0$\pm$0.8} \\
\hline
\multicolumn{5}{|c|}{\textit{GPT-3.5-Turbo-1106}} \\
\hline
Direct~\cite{soen}
    & 64.4$\pm$3.7 & 49.8$\pm$3.0 
    & 77.5$\pm$0.8 & 53.9$\pm$0.7 \\
CoT~\cite{islam2024mapcodermultiagentcodegeneration}
    & 68.9 & -- & 54.5 & -- \\
Self-Planning~\cite{shinn2023reflexionlanguageagentsverbal}
    & 60.3 & -- & 55.7 & -- \\
Reflexion~\cite{shinn2023reflexionlanguageagentsverbal}
    & 67.1 & -- & 73.0 & -- \\
LDB~\cite{zhong2024debuglikehumanlarge}
    & 82.9 & -- & 76.0 & -- \\
MapCoder~\cite{islam2024mapcodermultiagentcodegeneration}
    & 80.5 & 70.1 & 78.3 & 54.4 \\
\textbf{PyCapsule (ours)}
    & \textbf{85.2$\pm$0.7} 
    & \textbf{84.7$\pm$0.6} 
    & \textbf{78.5$\pm$1.2} 
    & \textbf{62.4$\pm$0.6} \\
\hline
\multicolumn{5}{|c|}{\textit{Qwen2.5-Coder-7B-Instruct}} \\
\hline
Direct~\cite{hui2024qwen25codertechnicalreport}
    & 88.4 & 84.1 
    & \textbf{83.5}$^{a}$ 
    & \textbf{71.7}$^{a}$ \\
\textbf{PyCapsule (ours)}
    & \textbf{94.1$\pm$1.3} 
    & \textbf{93.3$\pm$0.6} 
    & 80.7$\pm$0.9 
    & 63.6$\pm$0.6 \\
\hline
\multicolumn{5}{l}{}\\[-5pt]
\multicolumn{5}{@{}p{21pc}@{}}{\hspace*{9pt}$^{a}$Qwen2.5 employed 3-shot prompting for MBPP evaluation to mitigate dataset instability, as documented in their methodology. 
This represents a different experimental setting compared to the zero-shot approach used by us and other baselines.}
\end{tabular*}
\end{table}

\begin{table}[t]
\caption{PyCapsule Success Rate on BigCodeBench Compared to Direct Zero-Shot Baselines. 
$\Delta$ denotes the absolute improvement over the direct baseline.}
\label{table_bcb}
\tablefont%
\setlength{\tabcolsep}{3pt}
\begin{tabular}{|p{110pt}<{\centering}
                 |p{35pt}<{\centering}
                 |p{50pt}<{\centering}
                 |p{30pt}<{\centering}|}
\hline
\textbf{Model} & 
\textbf{Direct(\%)} & 
\textbf{PyCapsule(\%)} & 
$\boldsymbol{\Delta}$ \\
\hline
Qwen2.5-7B~\cite{hui2024qwen25codertechnicalreport}
    & 45.8 & \textbf{57.28} & +11.48  \\
Qwen2.5-Coder-Instruct-7B~\cite{hui2024qwen25codertechnicalreport}
    & 41.0 & \textbf{60.18} & +19.18 \\
Qwen2.5-Coder-Instruct-14B~\cite{hui2024qwen25codertechnicalreport}
    & 48.4 & \textbf{70.09} & +21.69 \\
Qwen2.5-Coder-Instruct-32B~\cite{hui2024qwen25codertechnicalreport}
    & 49.6 & \textbf{74.12} & +24.52 \\
Phi-4-14B~\cite{zhuo2024bigcodebench}
    & 50.5 & \textbf{68.07} & +17.57  \\
\hline
\end{tabular}
\end{table}



Table~\ref{table_acc} presents a systematic comparison of PyCapsule against established baselines across HumanEval, HumanEval-ET, MBPP, and MBPP-ET. 
PyCapsule achieves SOTA results on HumanEval and HumanEval-ET across all three model configurations, and competitive performance on MBPP and MBPP-ET. 
The sole exception arises in the Qwen2.5-Coder-Instruct configuration on MBPP variants, where the reported direct baseline employs three-shot prompting, a fundamentally different experimental setting from the zero-shot approach used consistently across all PyCapsule evaluations. 
To ensure comparability and reproducibility, we maintain zero-shot prompting throughout, employing the signature converter module to supply structured function signatures in place of few-shot exemplars. 
Within this consistent setting, PyCapsule with Qwen2.5-Coder-Instruct-7B achieves a 94.1\% success rate on HumanEval, surpassing not only the standalone zero-shot performance of the 32B variant but also the direct baseline of proprietary models such as GPT-4 at 78.86\% -- a result that motivates a broader evaluation across more demanding benchmarks. 

Table~\ref{table_bcb} extends this analysis to BigCodeBench, a significantly more challenging benchmark encompassing diverse real-world programming tasks involving complex function calls and library usage.
PyCapsule delivers substantial gains over direct zero-shot baselines regardless of model size or family. 
Notably, PyCapsule with Qwen2.5-Coder-Instruct-7B surpasses the standalone performance of the Qwen2.5-Coder-Instruct-32B model, a result that carries particular weight given that both models were released concurrently and trained on closely aligned data, making parameter count the principal differentiating factor. 
This demonstrates that PyCapsule can bridge a performance gap that would otherwise require a model more than four times the size, with direct implications for deployment in resource-constrained environments. 
The gains observed across the Qwen2.5-Coder family further suggest that PyCapsule's self-debugging mechanism scales effectively with model capacity.

Taken together, the consistent improvements across HumanEval and BigCodeBench -- benchmarks that differ substantially in problem complexity, domain coverage, and evaluation methodology demonstrate that PyCapsule generalises effectively across task distributions. 
The comparatively modest MBPP result, attributable to the methodological asymmetry discussed above, should be interpreted within that broader context rather than as a limitation of the framework.

\subsection{Influence of Self-debugging Attempts}
\begin{figure*}[h]
    \centering
    \includegraphics[width=\linewidth]{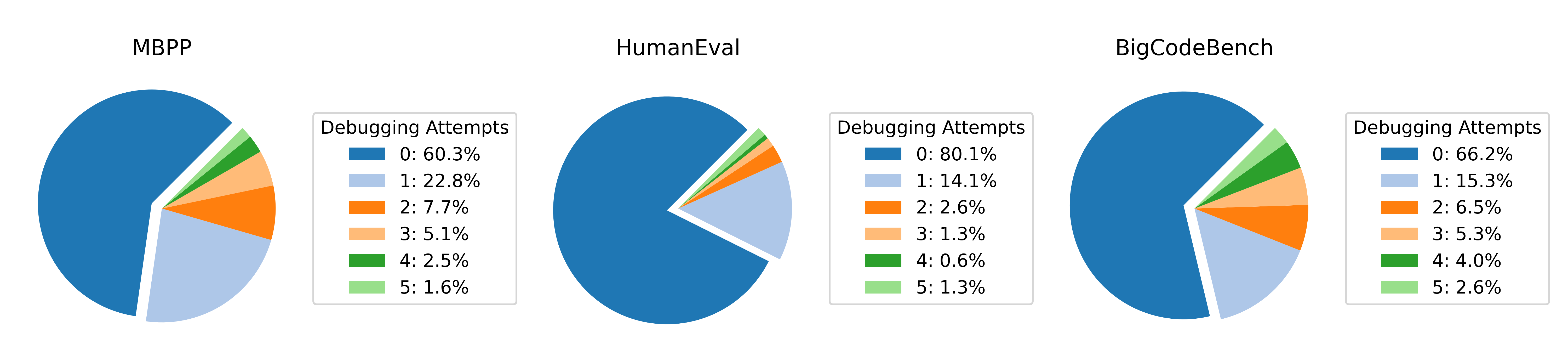}
    \caption{Distribution of relative success ratios along the self-debugging attempts.
    The relative success ratio refers to the number of successful test cases with given attempts over the total number of successful test cases, measured by unit \%.
    Results on all the datasets consistently show an exponential decrease in the relative success ratio as the number of attempts increases.}
    \label{fig:pie_fig}
\end{figure*}

To evaluate the contribution of each self-debugging attempt to the overall success rate, we analyse the normalised 
independent influence of each incremental attempt across GPT-4-1106-preview, GPT-3.5-Turbo-1106, and Qwen-2.5-Coder-Instruct 7B.
Given $N$ problems, the number of problems successfully solved at attempt zero is denoted as $S_0$, with the remaining $N_1 = N - S_0$ unsolved problems proceeding to the next debugging attempt.
The number of problems solved at attempt $i \geq 1$ is denoted as $S_i$, and $N_{i+1} = N_i - S_i$ represents the residual unsolved set.
The independent influence is defined as -- 
\begin{equation*}
    I_i = \frac{S_i}{N_i}, \text{where } N_i = N - \sum_{j=0}^{i-1} S_j
\end{equation*}

This metric isolates the contribution of each attempt, independent of cumulative progress.
Fitting the influence data to an exponential decay function yields $R^2$ values close to 1 across all models and datasets, confirming that debugging effectiveness decays exponentially with successive attempts. 
This pattern is consistent with prior findings on debugging effectiveness decay~\cite{ddi} and is here corroborated across multiple model families and scales. 
Notably, the rate of decay varies with model capability: GPT-4-Preview-1106 approaches negligible returns by the third attempt, whereas Qwen2.5-Coder-Instruct-7B sustains meaningful contributions through the fifth. 
This suggests that more capable models resolve residual errors more decisively in early attempts, leaving fewer tractable problems for later iterations. 
Accordingly, five debugging attempts were adopted as the architectural maximum, capturing the full extent of recoverable gains across all evaluated models whilst avoiding unnecessary computational overhead.

\begin{figure}[!h]
    \centering
    \includegraphics[width=\linewidth]{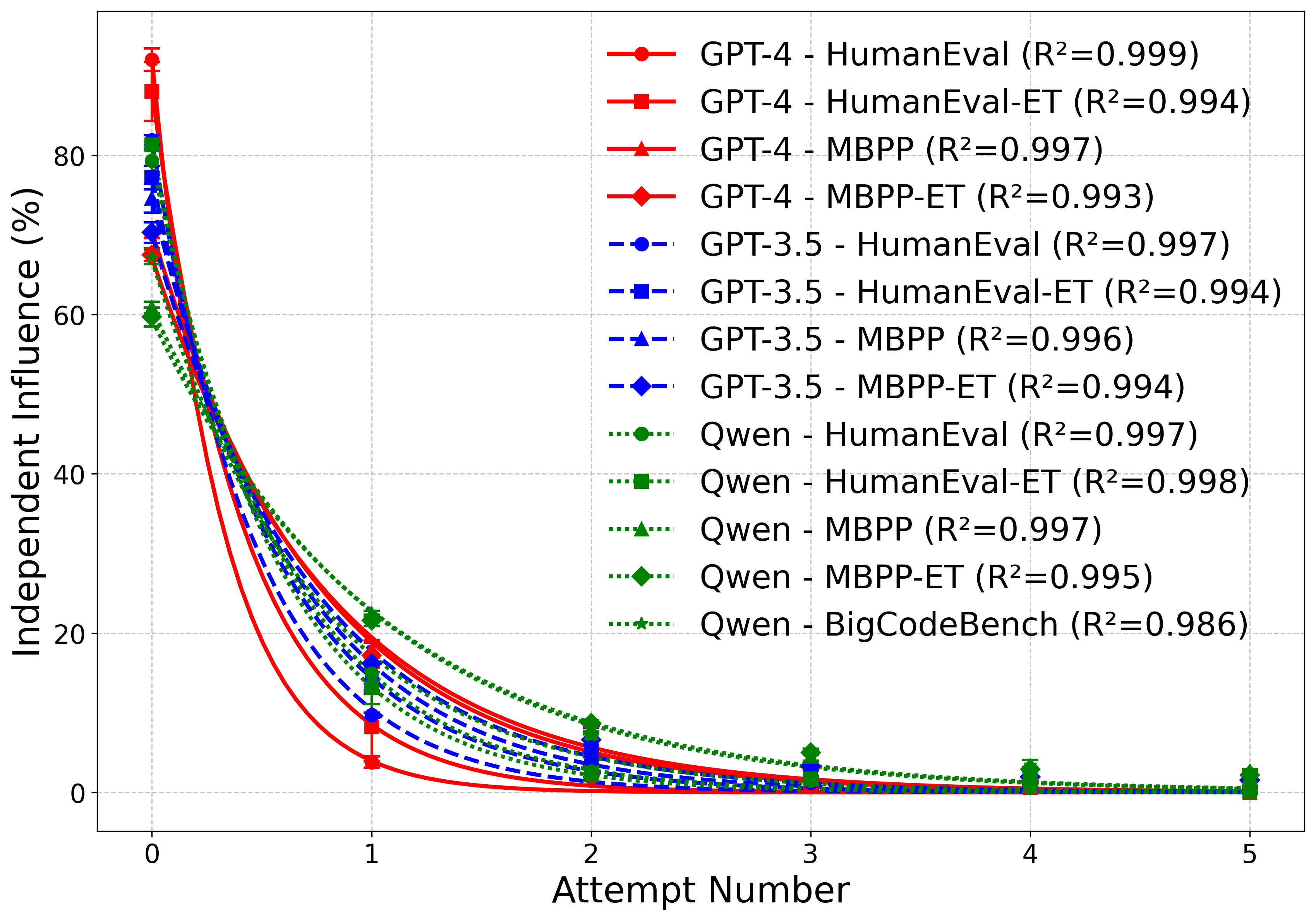}
    \caption{The plot illustrates the exponential decay in the normalised independent influence of self-debugging attempts across models. It shows the mean effectiveness of each debugging attempt, fitted to the exponential decay function $I(x) = a \cdot e^{-bx}$. The excellent fit of this function serves as an ideal guide for estimating the influence of successive attempts and can be used as a metric to compare the performance of different models.}
    \label{fig:influence}
\end{figure}

\section{Ablation Study}
\label{ablation}
To assess the independent contribution of each supporting module, we conduct an ablation study using Qwen2.5-Coder-7B-Instruct across HumanEval, MBPP, and BigCodeBench. 
This model was selected for three reasons. 
First, as a locally deployable open-source model, it enables fully reproducible ablation experiments that are not subject to the versioning uncertainty inherent in proprietary API-based systems. 
Second, being the smallest model in our evaluation, it is the most sensitive to the presence or absence of individual modules.
A more capable model may partially compensate for missing scaffolding, potentially obscuring the true contribution of each component. 
Third, repeated full-dataset runs are computationally feasible without incurring API costs, enabling the controlled, isolated evaluation that a rigorous ablation requires.

The \textit{signature converter} is employed exclusively for MBPP and MBPP-ET, where problem descriptions provide no structured function signature. 
To evaluate its contribution, we ran the full MBPP dataset without the module, yielding a success rate of 17.3\% compared to 80.7\% with the module active -- a difference of 63.4\%. 
Manual inspection of failure cases revealed a consistent and systematic failure mode. 
In the absence of a structured signature, the programmer agent generated functions with arbitrary names that did not match the identifiers expected by the test cases. 
Since MBPP test cases are expressed as assert statements that invoke a specific function name, any mismatch results in immediate failure regardless of the correctness of the underlying logic. 
This finding confirms that the signature converter is not merely a convenience feature but a prerequisite for stable code generation on datasets that lack structured function specifications.

The \textit{example call detector} addresses a distinct class of failure that does not manifest as a performance degradation in standard metrics but poses a meaningful risk to execution safety.
In our experiments, embedded example function calls were identified in 21 of 164 HumanEval problems and 44 of 344 MBPP problems when using GPT-3.5-Turbo-1106, despite explicit system prompts instructing the model to avoid them. 
These calls execute automatically during test case evaluation, bypassing the runtime safety mechanisms and producing uncontrolled side effects. 
In all flagged cases, the module successfully identified and removed the offending calls prior to execution, preventing runtime failures. 
On BigCodeBench, the module proved particularly critical: problems involving web application initialisation consistently triggered blocking operations, such as server startup routines embedded within generated code that caused execution to hang indefinitely until the maximum runtime limit was reached. 
The deterministic nature of this module ensures that such cases are handled reliably without requiring an additional LLM call for safety verification, which would substantially increase computational overhead.

To evaluate the contribution of \textit{the error-handling module}, we ran the full evaluation pipeline without it, passing raw executor output directly to the programmer agent. 
On HumanEval and MBPP, success rates declined by 9\% and 12\% respectively. 
On BigCodeBench, the decline was more modest at 4\%. This divergence is directly attributable to differences in error reporting structure across datasets. 
HumanEval and MBPP test cases are implemented as bare assert statements, which, upon failure, produce minimal diagnostic output -- typically a plain \texttt{AssertionError} with no contextual information. 
In this setting, the error handler's role in producing concise, contextually enriched natural language feedback is critical to guiding the programmer agent toward a meaningful fix. 
BigCodeBench, by contrast, employs structured Python unit tests with dedicated test classes and descriptive test method names, so the raw error output already carries sufficient diagnostic signal for the programmer agent to act on. 
The error handler, therefore, provides a smaller marginal gain on BigCodeBench, not because it is less effective, but because the baseline quality of unprocessed error messages is substantially higher in that setting. 
Taken together, these results confirm that the error handler provides substantial value when execution feedback lacks sufficient diagnostic information. 
By transforming sparse and often uninformative error messages into concise, contextually rich explanations, it enables the programmer agent to identify and correct failures more effectively. 
While the benefits are most pronounced in HumanEval and MBPP, where raw feedback is limited, the error handler remains beneficial across all benchmarks by improving the clarity and actionability of execution feedback.

\section{Discussion and Future Work}

\subsection{Prompt Engineering and First-Attempt Performance}
Figure~\ref{fig:pie_fig} and Figure~\ref{fig:influence} reveal that PyCapsule achieves substantially higher first-attempt accuracy than competing frameworks across all model configurations, even before self-debugging is applied. 
This establishes that PyCapsule's gains are not solely attributable to iterative correction but reflect a more fundamental advantage in code generation quality from the outset. 
The structured system prompts employed in generation mode, which define a clear programmer persona, enforce chain-of-thought reasoning, and provide explicit output formatting constraints reduce the likelihood of syntactically malformed or semantically misaligned responses at the first attempt. 
Prior work has demonstrated that prompt design critically affects model performance~\cite{Wei2022ChainOT, prompt_eg_consistency, eval_prompt_eg}, and our results corroborate this at scale across multiple model families. 
The subsequent self-debugging mechanism then compounds this advantage, as a higher-quality initial solution requires fewer corrective iterations to reach correctness, consistent with the exponential decay in debugging effectiveness reported in Section~\ref{result}.

\subsection{Computational Efficiency}
A central design objective of PyCapsule is to achieve high accuracy without incurring the computational overhead characteristic of multi-agent frameworks. 
Existing frameworks such as MapCoder require up to 17 API calls per problem on HumanEval and 12 on MBPP, scaling at $k \times t$ where $k$ is the number of retrieved exemplars and $t$ is the number of debugging attempts. 
PyCapsule eliminates retrieval entirely and consolidates planning and code generation into a single API call, capping the maximum at 6 calls per problem. 
In practice, the empirical average is substantially lower: 2.09, 1.38, and 1.55 API calls per problem on HumanEval for GPT-3.5-Turbo-1106, GPT-4-Preview-1106, and Qwen2.5-Coder-Instruct-7B respectively.

Table~\ref{table_tokens} provides a detailed breakdown of token consumption and average debugging attempts across all model and dataset configurations. 
Average attempt counts range from 1.35 to 3.31, remaining well below the architectural maximum. 
It should be noted that token counting methodologies vary across frameworks, making direct numerical comparison with external systems inadvisable; the figures reported here are intended to characterise PyCapsule's internal resource consumption.

\begin{table}[t]
\caption{Token Usage and Debugging Attempt Analysis Across GPT-3.5, GPT-4, and Qwen2.5-Coder-Instruct on Standard Code Generation Benchmarks. 
Token counts for GPT models were obtained using OpenAI's \texttt{tiktoken} library, and for Qwen2.5-Coder-Instruct using its native tokeniser.}
\label{table_tokens}
\tablefont%
\setlength{\tabcolsep}{3pt}
\begin{tabular}{|p{62pt}<{\centering}
|p{48pt}<{\centering}
|p{48pt}<{\centering}
|p{50pt}<{\centering}|}
\hline
\textbf{Dataset} &
\textbf{Total Tokens} &
\textbf{Avg. Tokens/ Problem} &
\textbf{Avg. Attempts/ Problem} \\
\hline
\multicolumn{4}{|c|}{\textit{GPT-3.5-Turbo-1106}} \\
\hline
HumanEval    & 69,502  & 423.79  & 2.05 \\
HumanEval-ET & 75,263  & 458.92  & 2.10 \\
MBPP         & 432,144 & 443.68  & 2.46 \\
MBPP-ET      & 671,316 & 689.24  & 3.22 \\
\hline
\multicolumn{4}{|c|}{\textit{GPT-4-Preview-1106}} \\
\hline
HumanEval    & 79,342    & 483.80   & 1.35 \\
HumanEval-ET & 82,214    & 501.30   & 1.37 \\
MBPP         & 779,440   & 800.25   & 2.03 \\
MBPP-ET      & 1,044,909 & 1,072.80 & 2.79 \\
\hline
\multicolumn{4}{|c|}{\textit{Qwen2.5-Coder-7B-Instruct}} \\
\hline
HumanEval    & 70,868    & 432.13   & 1.60 \\
HumanEval-ET & 74,151    & 452.14   & 1.61 \\
MBPP         & 646,635   & 663.90   & 2.54 \\
MBPP-ET      & 832,728   & 854.96   & 3.31 \\
BigCodeBench & 1,960,864 & 1,720.06 & 3.15 \\
\hline
\end{tabular}
\end{table}

\subsection{Limitations}
Whilst PyCapsule demonstrates strong generalisation across benchmark datasets, several limitations warrant consideration. 
First, as evidenced by the ablation study, certain modules are necessitated by structural properties of specific datasets rather than general characteristics of code generation tasks. 
Deploying PyCapsule on a new task distribution may require analogous dataset-specific interventions, introducing an engineering overhead that a more general-purpose signature inference mechanism could address in future work.
Second, the benchmark datasets used in this evaluation --- HumanEval, MBPP, and BigCodeBench feature well-defined function boundaries, deterministic test cases, and controlled problem scopes. 
Real-world software development tasks rarely conform to these conditions, involving ambiguous requirements, stateful systems, cross-file dependencies, and potentially incomplete test suites. 
PyCapsule's self-debugging mechanism relies on executable test feedback as its primary correction signal; in settings where such feedback is unavailable or unreliable, its iterative correction capability would be substantially diminished.

\subsection{Future Work}
Several directions present themselves for extending this work. 
First, developing a generalised signature inference mechanism that operates without dataset-specific configuration would substantially improve PyCapsule's applicability to novel task distributions. 
Second, evaluating PyCapsule on real-world software engineering tasks, such as repository-level code completion or issue resolution would provide a more direct assessment of its practical utility beyond curated benchmarks. 
Finally, extending PyCapsule to support multi-language code generation would broaden its applicability, given the current Python-only focus of the framework.

\section{Conclusion}
\label{con}
We propose PyCapsule for significantly advancing reliable and automated code generation by integrating two AI agents with high-efficiency self-debugging modules. 
Its two-agent architecture effectively balances code generation automation and control while reducing resource consumption. Experimental results show that PyCapsule achieves high performance across multiple datasets, surpassing existing debugging-based approaches. 
Notably, PyCapsule enhances accuracy while using smaller language models and fewer API calls, offering a cost-effective alternative without compromising its performance. The iterative self-debugging feedback and dynamic prompt augmentation contribute to its robustness and improvements.

\bibliographystyle{IEEEtran}
\bibliography{references}

\appendix
\subsection{System Prompts}
\label{appendix: prompt}
\subsubsection{Code Generation Mode}
You are an experienced Python developer. 
Your task is to complete the function based on the provided function signature and description.
\begin{itemize}
    \item Analyse the description and provide a step-by-step reasoning on how to solve the problem.
    \item Maintain the function signature: Do not alter/modify the function signature.
    \item Avoid example calls. Do not include any example call in your response.
    \item If external libraries are needed, add a '\#\#\# Requirements' Section listing them separated by ','. e.g. pandas, python-dotenv.
    If no external libraries are needed, add 'None'.
    \item Import all required libraries and complete the function in a '\#\#\# Code' Section, enclosed with triple backticks.
\end{itemize}
Example Response Structure-\\
\#\#\# Step-by-step reasoning\\
\$reasoning\\
\\
\#\#\# Requirements\\
\$external\_libraries\\
\\
\#\#\# Code\\
\$code\\

\subsubsection{Fix Mode}
You are an experienced Python developer. 
\begin{itemize}
    \item Your previous solution resulted an error.
    \item Error message from a python compiler and Conversation history has been added for your reference.
    \item Analyse the description and provide a step-by-step reasoning on how to solve the problem.
    \item Maintain the function signature: Do not alter/modify the function signature.
    \item Avoid example calls: Do not include any example call in your response.
    \item If external libraries are needed, add a '\#\#\# Requirements' Section listing them separated by ','. e.g. pandas, python-dotenv.
    If no external libraries are needed, add 'None'.
    \item Import all required libraries and complete the function in a '\#\#\# Code' Section, enclosed with triple backticks.
\end{itemize}

Example Response Structure-\\
\#\#\# Step-by-step reasoning\\
\$reasoning\\
\\
\#\#\# Requirements\\
\$external\_libraries\\
\\
\#\#\# Code\\
\$code

\subsection{Pseudo-code of PyCapsule}
\label{appendix: pseudo}
\footnotesize
\begin{algorithm}[!h]
\caption{Pseudo-code of PyCapsule}\label{algo2}
\begin{flushleft}
\small
\textbf{Input:} A user-defined query $\mathcal{Q}$ for code generation with a problem description, test cases, and the maximum number of self-debugging attempts $N=5$.\\
\textbf{Output:} A generated code satisfying the request in the problem description if all test cases passed, otherwise a failed status message.\\
\textbf{Procedure:}\\
\end{flushleft}
\small
\begin{algorithmic}[1]
\State Set and initialise a self-debugging attempt counter $i=0$.
\State Set an initial system prompt for generating code, a buffer $\mathcal{B}$ for storing conversation history, and a user prompt initialised by $\mathcal{Q}$.
\State Use $\mathcal{Q}$ and the LLM in PyCapsule to generate the function signature required by case testing.
\State \textbf{Step 1:} Generate code (``Generation Mode").
\State Generate a prompt by integrating the system prompt and user prompt.
\State Use the prompt and the same LLM to generate a response for code generation.
\State Extract the required code, that is the function implementation, from the response.
\State Create a Python container with the required packages installed. If it exists, skip this step.
\State Test all the cases in the container using the extract code, and return the code execute status.
\small
\If{\textit{all the test cases passed}}
  \State Return the extracted code as the optimal generated code.
\ElsIf {$i > N$}
  \State Return the failed message extracted from the code execute status.
\Else
  \State \textbf{Step 2:} Code Debugging (``Fix Mode").
  \State Extract the error message from the code execute status and add it to $\mathcal{B}$.
  \State Generate an error-handling prompt using the error-handling module.
  \State Update the system prompt for code debugging.
  \State Update the user prompt by integrating the error-handling prompt and messages in $\mathcal{B}$.
  \State Update $n = n+1$.
  \State Repeat \textbf{Step 1}.
\EndIf
\end{algorithmic}
\end{algorithm}

\end{document}